\documentclass[fleqn,usenatbib,useAMS]{mnras}
\usepackage{newtxtext,newtxmath}
\usepackage[T1]{fontenc}
\usepackage{ae,aecompl}

\usepackage[varg]{txfonts}
\usepackage{natbib}
\usepackage{booktabs}

\makeatletter
    \if@twoside
        \hypersetup{pdfpagelayout=TwoPageLeft, linkcolor=blue}
        \fi
\makeatother

\usepackage{ulem}
\usepackage{graphicx}	

\title{Positive feedback, quenching and sequential super star cluster (SSC) formation in NGC~4945}

\author[E. Bellocchi et al.]{
E. Bellocchi,$^{1}$\thanks{E-mail: enrica.bellocchi@gmail.com, enricbel@ucm.es. \newline Current address: Departamento de F\'isica de la Tierra y Astrofísica, Universidad Complutense de Madrid, 28040 Madrid, Spain.}
J. Martín-Pintado,$^{1}$
F. Rico-Villas,$^{1}$
S. Mart\'in$^{2,3}$
and I. Jim\'enez-Sierra$^{1}$
\\
$^1$ Centro de Astrobiolog\'ia (CAB), CSIC-INTA, Ctra de Torrej\'on a Ajalvir, km 4, 28850 Torrej\'on de Ardoz, Spain \\        
$^2$ European Southern Observatory, Alonso de C\'ordova, 3107, Vitacura, Santiago, 763-0355, Chile\\ 
$^3$ Joint ALMA Observatory, Alonso de Córdova, 3107, Vitacura, Santiago, 763-0355, Chile\\
}

\date{Accepted 24 November 2022; Received 1 November 2022}
\pubyear{2022}

\begin{document}
\pagerange{\pageref{firstpage}--\pageref{lastpage}}

\maketitle

\begin{abstract}
We have used ALMA imaging (resolutions 0.1\arcsec-0.4\arcsec) of ground and vibrationally excited lines of HCN and HC$_3$N toward the nucleus of NGC~4945 to trace the protostellar phase in Super Star Clusters (proto-SSC). Out of the 14 identified SSCs, we find that 8 are in the proto-SSC phase showing vibrational HCN emission with 5 of them also showing vibrational HC$_3$N emission. We estimate proto-SSC ages of 5-9.7$\times$10$^4$ yr. The more evolved ones, with only HCN emission, are close to reach the Zero Age Main Sequence (ZAMS; ages $\gtrsim$10$^5$ yr). The excitation of the parental cloud seems to be related to the SSC evolutionary stage, with high ($\sim$65 K) and low ($\sim$25 K) rotational temperatures for the youngest proto and ZAMS SSCs, respectively. Heating by the HII regions in the SSC ZAMS phase seems to be rather local. The youngest proto-SSCs are located at the edges of the molecular outflow, indicating SSC formation by positive feedback in the shocked regions. The proto-SSCs in NGC~4945 seem to be more evolved than in the starburst galaxy NGC~253. We propose that sequential SSC formation can explain the spatial distribution and different ages of the SSCs in both galaxies.
\end{abstract}


\vspace{-10mm}
\section{Introduction}

Bursts of star formation in galaxies are usually concentrated in small regions know as Super Star Cluster (SSC), which are compact in size ($\sim$1 pc), massive ($M_{\rm \star}$$\sim$10$^5$ $M_{\sun}$) and young (from 1 up to 100 Myr). 
This mode of star formation is thought to play a major role in the evolution of galaxies, specially since it seems to be more frequent in galaxies with high star formation rates resulting from merging events, as in starbursts (SB) and in luminous and ultraluminous systems.
It is expected that in a very short time scales (few 10$^{6-7}$ yr) the most massive stars in SSCs will explode as supernova ejecting hot shells that eventually will generate a powerful outflow. So far, it is unclear to what extent the evolution of the SSC will produce a sequential SSCs formation due to positive feedback like in the Antennae galaxies (\citealt{Herrera11}) or whether it will quench star formation (\citealt{Wall20}). 
The study of SSCs in nearby galaxies is key in order to understand the triggering mechanisms that lead to the starbursts, and to establish the effects of the feedback processes (positive versus negative) in the earliest stage of their evolution. 
However, the earliest phases of SSCs formation take place deeply embedded in the parental cloud, preventing their observation even at mid-IR.
Recently, \cite{Rico20} have identified and studied the formation and the earliest stages of  evolution of the SSCs, deeply embedded in the parent clouds, in the starburst galaxy NGC~253 by means of the HC$_3$N vibrationally excited (HC$_3$N$^*$) emission. They found that the SSCs in NGC~253 seem to be formed by the overpressure generated by the hot expanding gas produced by the SN explosion(s) (timescale $\sim$1~Myr) from an earlier star formation episode in the galaxy center but unrelated to the outflow, thus suggesting sequential SSC formation. 

NGC~4945, the closest (D$\sim$3.8 Mpc) composite galaxy in which an AGN and SB co-exist, shares with NGC~253 similar properties and they are both located at similar distances. 
Both galaxies have a star formation rate (SFR) of about 3 $M_{\sun}$ yr$^{-1}$ and show the presence of an outflow both in X-rays, optical and in the molecular emissions (see \citealt{Schurch02, Mingozzi19, Bellocchi20, Bolatto21} for NGC~4945 and \citealt{Strickland02, Westmoquette11, Bolatto13, Krieger19} for NGC~253): their ionized outflows share similar momentum flux and mass outflow rate (\citealt{Heckman90}). 
However, the molecular outflow in NGC~4945 seems less massive (\citealt{Krieger19}) and slightly less powerful (\citealt{Bolatto21}) than that in NGC~253.
The main difference between both galaxies is the age of the SB: in NGC~253 it is relatively younger ($\sim$1 Myr; \citealt{Leroy18}) than that in NGC~4945 ($\sim$5 Myr; \citealt{Emig20}). Then the study of the formation and early evolution of the SSCs in NGC~4945, when compared to the one already carried out toward NGC~253 (\citealt{Rico20, Rico21,Rico22}), has the potential to provide further information on the evolution of the starburst and to establish to what extent the sequential star formation (positive feedback) also dominates at later stages or whether the SSC formation has been quenched.  

In this work we use ALMA data to study the early evolutionary stage of the proto-SSCs using the HCN and HC$_3$N vibrationally excited emission in NGC~4945.
From  the location and the ages of the young SSCs we found that the proto-SSCs are older than in NGC~253 and that the last generation of SSCs in NGC~4945 have been triggered by the positive feedback of the molecular outflow (\citealt{Bolatto21}) arising from the previous generations of SSCs.

\begin{table}
\centering
\begin{small}
\caption{Coordinates and summary of the detected HC$_3$N lines of the selected SSCs in NGC~4945. }
\label{Input_regions}
\begin{tabular}{lccccc} 
\hline
\multicolumn{1}{c}{SSC}	 &		RA & 	DEC	 & Emig \# & \multicolumn{2}{c}{HC$_3$N$^*$}\\
  & 13$^h$05$^m$       &-49$^\circ$28$^\prime$   &  &    $v_{\rm 71}$, $v_{\rm 61}$ & $v_{\rm 72}$    \\
\hline
proto-HCN-1 & 27$^s$.751  & 01\arcsec.89	& 2  &   u & u  \\
proto-HC$_3$N-2 & 27$^s$.540  & 03\arcsec.99	& 12 &  	 $\checkmark$	& $\checkmark$  \\
proto-HC$_3$N-3 & 27$^s$.530  & 04\arcsec.19 &13 &  	    $\checkmark$	&$\checkmark$ \\
proto-HC$_3$N-4 & 27$^s$.505  & 04\arcsec.04	& 16 	&  $\checkmark$	& $\checkmark$  \\
proto-HCN-5 & 27$^s$.448 & 04\arcsec.69	& \dots &  	 u & u  \\
proto-HC$_3$N-6 & 27$^s$.402 & 07\arcsec.14	& 21 &     $\checkmark$ & u \\
proto-HCN-7 & 27$^s$.361 & 06\arcsec.04	& 22 &   u &  u  \\
proto-HC$_3$N-8 & 27$^s$.284 & 06\arcsec.69	& 26 & $\checkmark$	& u  \\
\hline
ZAMS-9 & 27$^s$.590 & 03\arcsec.44 & 8  & \dots & \dots \\
ZAMS-10 &27$^s$.489 & 05\arcsec.14   & 17  & \dots & \dots \\
ZAMS-11 & 27$^s$.457 & 05\arcsec.76  & 20  & \dots & \dots \\
ZAMS-12 & 27$^s$.288 & 06\arcsec.56  & 25  & \dots & \dots \\
ZAMS-13 & 27$^s$.269 & 06\arcsec.60  & 27 & \dots & \dots \\
ZAMS-14 &27$^s$.197 & 08\arcsec.19   &   29  & \dots & \dots \\
\hline
\end{tabular}
\end{small}
\begin{minipage}{8.cm}
{{\bf Notes}: Column~(4): Region number from \cite{Emig20}. Columns~(5-6): Detection ($\checkmark$) or upper limit ({\it u}) of the HC$_3$N$^*$ emission.}
\end{minipage}
\end{table}

\vspace{-0.5cm}
\section{Data reduction}
\label{obs_data}

We obtained the HCN, HC$_3$N and CS images toward NGC 4945 from the ALMA Science Archive data (Project-IDs: 2016.1.01135.S and 2018.1.01236.S). 
For our analysis we have extracted the data for the following transitions at $\sim$100, $\sim$340 and $\sim$355 GHz: HCN(4-3) from the vibrational excited level $v_{\rm 2}$=1 (hereafter, HCN$^*$), and the HC$_3$N(11-10), (38-37) and (39-38) from the ground (hereafter, HC$_3$N,v$_0$) and vibrationally excited (hereafter, HC$_3$N$^*$) levels $v_{\rm 7}$=1, $v_{\rm 6}$=1 and $v_{\rm 7}$=2 (hereafter, $v_{\rm 71}$, $v_{\rm 61}$, $v_{\rm 72}$, respectively). The CS(2-1) and CS(7-6)  lines were occasionally used when only one HC$_3$N transition was detected.

The data were reduced using the Common Astronomy Software Applications (CASA, McMullin et al 2007) version 4.2.2. The CASA's {\tt clean} task with Briggs weighting was used for the deconvolution, setting the {\tt robust} parameter to 0.5. The velocity resolution of the data cubes was smoothed to 8~km~s$^{-1}$. Primary beam correction was applied. The final synthetized beams range between 0.14\arcsec$\times$0.11\arcsec\ ($\sim$100 GHz) to 0.42\arcsec$\times$0.37\arcsec\ ($\sim$350 GHz) and the sensitivity, in 8~km~s$^{-1}$, varies between 0.3 mJy ($\sim$100 GHz) and 0.9 mJy ($\sim$350 GHz). Throughout the paper, we adopted a systemic velocity of 550~km~s$^{-1}$, based on CO observations (\citealt{Dahlem93}).

 \begin{figure}
 \centering
 \includegraphics[width=0.37\textwidth]{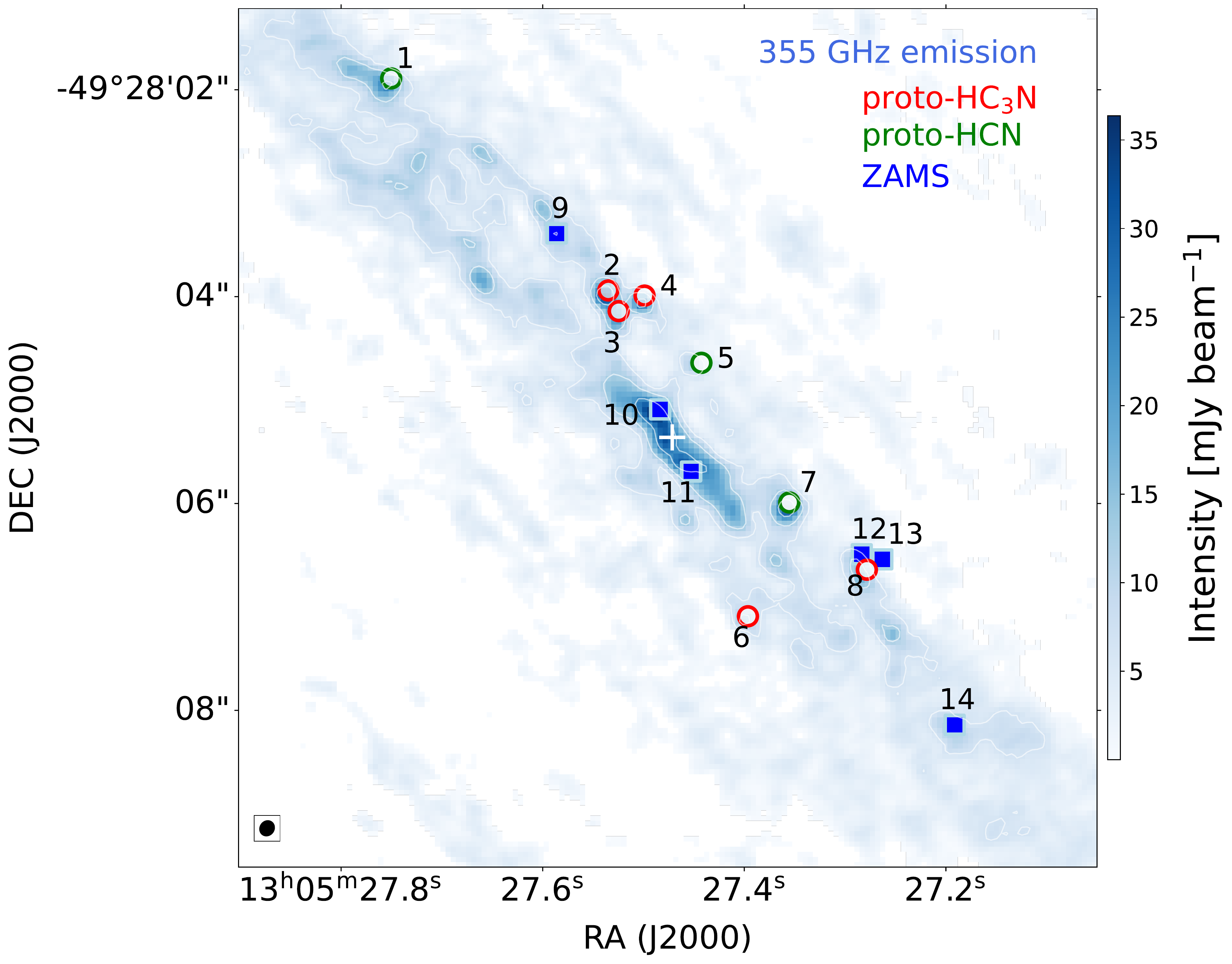}
 \caption{
 Dust emission at $\sim$350 GHz with contours overlaid. The nucleus is identified by the white cross. The red and green open circles show the proto-SSCs where HCN$^*$ is detected: among these, those also showing HC$_3$N,$v_{\rm 71}$ emission are shown in red open circles (proto-HC$_3$N) while those without detection of the HC$_3$N,$v_{\rm 71}$ emission are shown in green open circles (proto-HCN). The filled square in blue have been selected from the dust peak emissions without detection of HCN$^*$ or HC$_3$N$^*$ (ZAMS-SSCs). The beam size is shown in the lower left corner. The SSCs are ordered according to the positions described in Table~\ref{Input_regions}. 
 }
\label{positions_regs}
\end{figure}

\vspace{-0.5cm}
\section{Super Star Cluster (SSC) selection}

We have used the vibrationally emission HCN$^*$ and HC$_3$N$^*$ to trace the Super Star Clusters (SSCs) in their earliest phases of evolution (\citealt{Rico20, Rico21}), known as the Super Hot Core (SHC) phase. 
We selected 14 SSCs, out of 27 found by \cite{Emig20}, in two ways: i) by the peak emissions of the vibrationally excited HC$_3$N,v$_{71}$ or the HCN$^*$ and, ii) by the dust continuum peak at $\sim$350 GHz with only CS emissions. The ground state peak emissions of HC$_3$N,v$_0$ is observed in all the 14 SSCs. 
In the latter case, we searched for the SSCs characterized by high stellar mass (10$^{5.1}$-10$^{5.7}$ $M_{\sun}$) as measured by the radio continuum emission together with high ($>$10$^{5.0}$~$M_{\sun}$) or low ($<$10$^{5.0}$~$M_{\sun}$) gas mass (\citealt{Emig20}). 
This selection allows to study their evolutionary stage. Fig~\ref{positions_regs} shows the locations of the selected SSCs superimposed on the 345 GHz continuum emission and Table~\ref{Input_regions} gives the positions and a summary of the detected lines.
Among all the selected regions, HCN$^*$ emission has been found in 8 (proto-SSCs) of the 14 SSCs, which are then considered as putative SHCs in the proto-SSC phase. 
In fact, among them, we have also detected HC$_3$N$^*$ in 5 proto-SSCs (hereafter, proto-HC$_3$N) while for the remaining three we derived an upper limit to this emission (hereafter, proto-HCN; see Table~\ref{Input_regions}). 
In most of the cases the positions derived from the HC$_3$N$^*$ and HCN$^*$ emissions are in good agreement with the condensations observed in the dust maps within the beam  (see Fig.~\ref{positions_regs}).
The remaining 6 dust selected regions, hereafter ZAMS-SSCs, do not show the SHC phase since neither HCN$^*$ nor HC$_3$N$^*$ emissions have been detected.


\vspace{-0.7cm}
\section{Analysis}
\label{sect4}

\begin{figure}
 \centering
\includegraphics[width=0.47\textwidth]{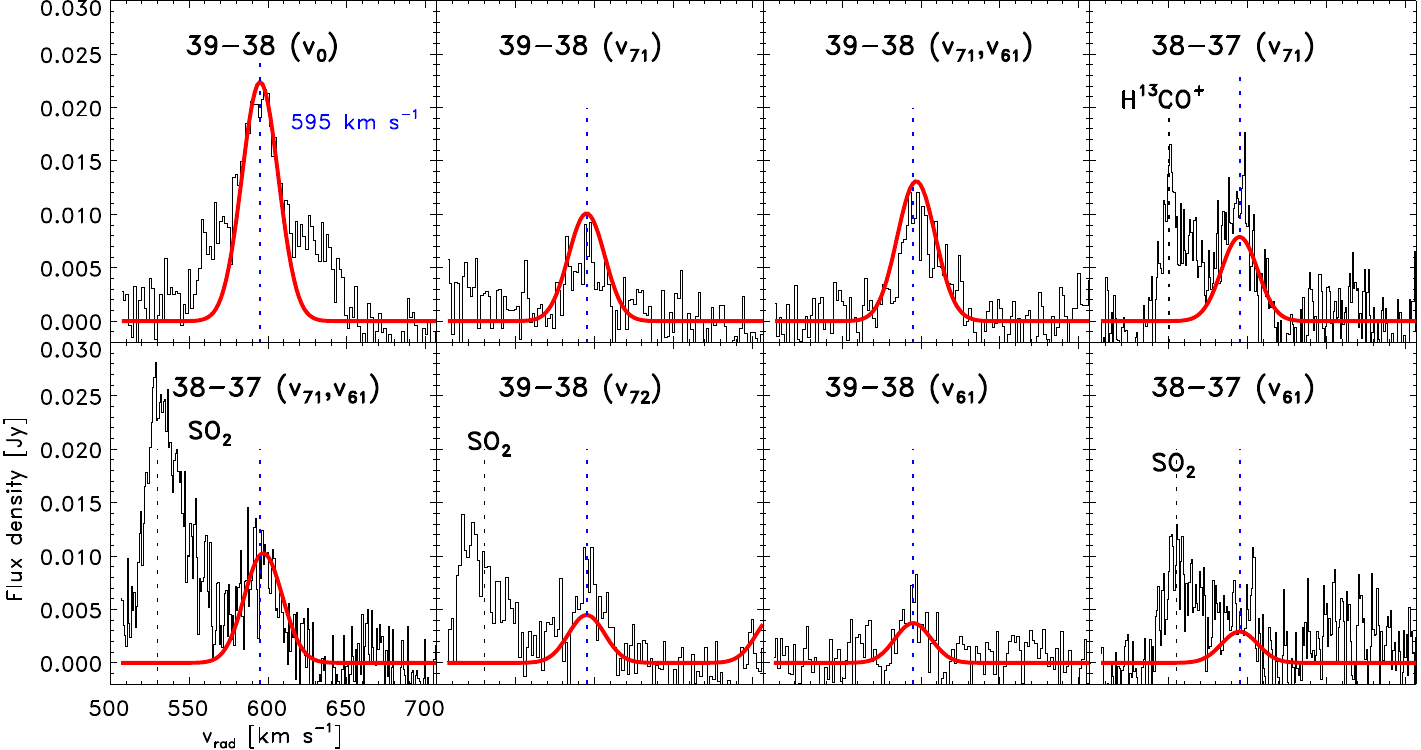}
\vskip-2mm\caption{Observed HC$_3$N emission at $\sim$350 GHz (black histograms) of the brightest proto-HC$_3$N-4. The red solid lines represent the fitted model from the LTE analysis carried out with MADCUBA. The transitions from the ground J=39-38 ($v_{\rm 0}$) and vibrationally excited states J=39-38 and J=38-37 ($v_{\rm 71}$ and $v_{\rm 61}$) are shown on the top of each panel. Emission lines found near the HC$_3$N$^*$ emissions are also identified. The systemic velocity of the SSC is shown in blue by a vertical dotted line.}
 \label{119112_spectra}
 \end{figure}

We used the MADCUBA\footnote{Madrid Data Cube Analysis (MADCUBA) is a software developed in the Center for Astrobiology in Madrid able to visualize and analyze data cubes and single spectra under Local Thermodynamic Equilibrium (LTE) conditions. See \url{https://cab.inta-csic.es/madcuba/}}'s tool SLIM (Spectral Line Identification and Modelling; \citealt{Martin19}) to identify the molecular lines and to perform the Local Thermodynamic Equilibrium (LTE) analysis. This tool uses the molecular spectroscopy in the publicly available molecular catalogues CDMS\footnote{\url{https://cdms.astro.uni-koeln.de/cgi-bin/cdmssearch}} (\citealt{Muller01}) and JPL\footnote{\url{https://spec.jpl.nasa.gov/ftp/pub/catalog/catform.html}} (\citealt{Pickett98}). 
SLIM simulates the line profiles under LTE conditions, including opacity effects for a given source size.
The AUTOFIT tool in SLIM performs a non-linear least squared fit of the LTE line profiles to the data using Levenberg-Marquartd algorithm. 
Fig.~\ref{119112_spectra} shows an example of the HC$_3$N spectra in the ground and vibrationally excited states together with the SLIM fit toward the brightest SHC (proto-HC$_3$N-4) in HC$_3$N$^*$.

Following the procedure depicted by \cite{Rico20}, we first focus on the determination of the physical parameters (e.g., excitation temperature, column density, radial velocity, line width and source size) of the proto-HC$_3$N in the SHC phase.
We derive the excitation temperature of the vibrationally excited levels (hereafter, $T_{\rm vib}$) by only considered transitions with the same J but arising from different vibrationally excited states. Since HC$_3$N$^*$ transitions are excited through the absorption of mid-IR photons and, at these wavelengths the dust emission is optically thick, the vibrational temperature is expected to closely reflect the dust temperature ($T_{\rm vib}$$\sim$$T_{\rm dust}$). 
To derive the excitation temperature of the rotational levels within the vibrational states (hereafter, $T_{\rm rot}$) we combined all the rotational transitions arising from the same vibrationally states (ground state, $v_{\rm 71}$, $v_{\rm 61}$ and $v_{\rm 72}$). 
For the proto-HCN, only traced by the HCN$^*$ lines, we can only derive an upper limit to $T_{\rm vib}$ ($\sim$$T_{\rm dust}$) from the LTE analysis of HC$_3$N J=11-10 (i.e., 100 GHz; see Table~\ref{Input_data}).
For all the SSCs in our sample, including the ZAMS-SSCs, we have also derived the $T_{\rm rot}$ from CS and HC$_3$N,v$_0$, which provides the typical excitation temperature of the bulk of the gas associated with the SSCs in this stage of evolution.

\begin{table}
\centering
\begin{small}
\caption{Results from the LTE analysis of the HC$_3$N and HC$_3$N$^*$ emissions.}
\label{Input_data}
\begin{tabular}{lcccc} 
\hline
\multicolumn{1}{c}{SSC}	 &	 $T_{\rm rot}$($v_{\rm 0}$) & $T_{\rm rot}$($v_{\rm 61}$)  & $T_{\rm rot}$($v_{\rm 71}$) & $T_{\rm vib}$  \\
 \hline
proto-HCN-1 &  $<$55 & \dots & \dots & $<$141   \\ 
proto-HC$_3$N-2 &   61 (1) & 76 (23) & 81 (14) & 335 (33)  \\
proto-HC$_3$N-3 &   66 (2) & 82 (89) & 90 (35) & 201 (15)   \\
proto-HC$_3$N-4 &   70 (4) & 95 (63) & 77 (10) & 398 (26)  \\
proto-HCN-5   &  $<$42 & \dots & \dots & $<$153 \\ 
proto-HC$_3$N-6 &  67 (14) & 79 (77) & 77 (16) & 197 (31) \\ 
proto-HCN-7 &  $<$43 & \dots & \dots & $<$132 \\ 
proto-HC$_3$N-8  &  58 (2) & 73 & 72 (15) & 412 (72)  \\ 
\hline
ZAMS-9  & 19.0$^\star$ & \dots & \dots & $<$159 \\
ZAMS-10  & 16.4$^\star$ & \dots & \dots & $<$117  \\
ZAMS-11  & 13.0$^\star$ & \dots & \dots & $<$99  \\
ZAMS-12  & 54.4 & \dots & \dots & $<$199  \\
ZAMS-13  & 46.3 & \dots & \dots & $<$126  \\
ZAMS-14  & 15.7$^\star$ & \dots & \dots & $<$176  \\
\hline
\end{tabular}
\end{small}
\hskip-10mm
\begin{minipage}{8.7cm}
\small
{{\bf Notes}: The ($^\star$) indicates that $T_{\rm rot}$ for the HC$_3$N fit has been fixed to that derived from CS. $T_{\rm vib}$ has been derived from the fit of the HC$_3$N$^*$ emission at 350 GHz (Sect.~\ref{sect4}). Upper limits to $T_{\rm vib}$ are derived from the 100 GHz LTE analysis. 
}
\end{minipage}
\end{table}

\section{Properties of the SSCs}

\subsection{Size of the proto-SSCs}
\label{size_par}

We estimated a lower limit to the proto-HC$_3$N size from the line intensities of the rotational transition from the $v_{\rm 71}$ by assuming that their emission is optically thick. 
When HC$_3$N$^*$ was not detected (i.e., proto-HCN and ZAMS-SSCs), we derived an upper limit to the size using the line intensity of the HC$_3$N,v$_0$, assuming the $T_{\rm vib}$ ($\sim$$T_{\rm dust}$) upper limit in Table~\ref{Input_data}. The derived the lower  and upper limits to the source size are shown in Table~\ref{Params_physical}.

\vspace{-0.5cm}
\subsection{Luminosities and masses}
\label{lum_derive}

We estimated the apparent luminosity of the proto-SSCs, $L_{\rm app}$, from the $T_{\rm vib}$ ($\sim$$T_{\rm dust}$; Table~\ref{Input_data}) and the lower (upper) sizes (Table~\ref{Params_physical}) using the Stefan-Boltzmann law: $L_{\rm app}$= 4$\pi$$r^2$ $\sigma$ $T_{\rm vib}^4$. 
These apparent luminosities of the proto-SSCs are, in principle, lower limits to the actual luminosity since we have used a lower limit to their sizes.
However, the luminosity derived in such a way only represents the actual luminosity in the case of low H$_2$ column densities ($N$($H_{\rm 2}$)~$<$10$^{23}$~cm$^{-2}$).
For larger column densities, as in our case, the luminosity needs to be corrected for the so-called `back-warming' or `green-house' effect (\citealt{GA19}).
Our apparent luminosity can be corrected of the green-house effect applying the correction factor 4/$\psi$ to estimate the actual luminosity of proto-stars ({$L_{\rm p*}$)} as {$L_{\rm p*}$}~=~4/$\psi$~$\times$ {$L_{\rm app}$}.
Considering a representative value of the $N$($H_{\rm 2}$)~of a few 10$^{24}$ cm$^{-2}$, and a typical density profile of the SHC with a power law index lying between 1 and 2, we estimated a correction factor 4/$\psi$$\sim$1/10 (see \citealt{Rico20}). Corrected (actual) luminosities are reported in Table~\ref{Params_physical}. 
Assuming the same light-to-mass ratio of 10$^3$ $L_{\sun}$ $M_{\sun}$$^{-1}$ used by \cite{Leroy18} to the derive the mass in stars in the ZAMS-SSCs, we can also estimate the mass in protostars, $M_{\rm p*}$ of the SSCs.

\vspace{-0.6cm}
\section{Discussion}

\subsection{HCN$^*$ emission. Tracing the latest stages of the proto-SSC phase before reaching the ZAMS-SSC stage. }

\cite{Rico20} established the evolutionary stage of the proto-SSCs in NGC~253 using the ratio between the $L_{\rm p*}$ and the $L_{\rm ZAMS}$. The derived ages for the proto-HC$_3$N using Eq.~3 of their work are given in Table~\ref{Params_physical}: we found that most of the proto-HC$_3$N have ages of 5-9.6$\times$10$^4$ yr ($\pm$1.5$\times$10$^4$ yr), close to reach the ZAMS stage with ages, $t_{\rm age}$, $\gtrsim$10$^5$~yr (\citealt{Rico20}). 
Our data also suggest that the proto-HCN are still in the proto-SSC phase since they show HCN$^*$ emission, typical of the SSC phase, without HC$_3$N$^*$ emission counterpart. In fact, the $L_{\rm p*}$/$L_{\rm ZAMS}$ ratio reach values as low as 0.03 (proto-HCN-7), indicating that at this stage, the fraction of protostar  mass in the SSCs is extremely small (Table~\ref{Params_physical}). The detection of proto-SSC through HCN allows to trace the proto-SSCs phase to much later stages than in NGC~253 as traced only by HC$_3$N$^*$ emission. We can then extent this method to estimate the ages of the SSCs by using the vibrationally excited HCN emission. The derived ages for the proto-SSCs-HCN are close to reach the ZAMS phase with an age close to 10$^5$~yr (Table~\ref{Params_physical}).

\vspace{-0.5cm}
 \subsection{The effect of the SSC evolution on the excitation of the parent cloud}

\begin{table}
\vspace{-0.3cm}
\begin{small}
\caption{SSC properties. }
\label{Params_physical}
\hskip-9mm
\begin{tabular}{l c @{\hskip3pt}c@{\hskip3pt}@{\hskip3pt}c@{\hskip3pt}@{\hskip3pt}c@{\hskip3pt}c@{\hskip3pt}c@{\hskip3pt}} 
\hline
  \multicolumn{1}{c}{SSC}  & Size &  $M_{\rm gas}$ & $L_{\rm p*}$ & $L_{\rm ZAMS}$ &    $L_{\rm p*}$/$L_{\rm ZAMS}$  &  $t_{\rm age}$  \\
 &	(pc) &	(10$^4$ $M_{\sun}$) &(10$^8$ $L_{\sun}$) & (10$^8$ $L_{\sun}$)  &  &(10$^4$ yr) \\
 \hline
p-HCN-1 & $<$0.76 & 6.3 & $<$0.1 & 1.58 & $<$0.06  & $>$9.4\\
p-HC$_3$N-2 & $>$0.28  & 10.0 & $\sim$0.42 & 1.58 & $\sim$0.27 &7.9 \\
p-HC$_3$N-3 & $>$0.41  & 5.0  & $\sim$0.12 & 3.16 & $\sim$0.04   & 9.6 \\
p-HC$_3$N-4 & $>$0.35  & 5.0  & $\sim$1.32 & 1.26 & $\sim$1.05   & 4.9\\
p-HCN-5 & $<$0.57 & 4.9$^{\star}$  & $<$0.08 & 1.26$^{\star\star}$ &  $<$0.06  & $>$9.4 \\
p-HC$_3$N-6 & $>$0.36 &  4.0 & $\sim$0.84 & 0.20 & $\sim$0.42  & 7.0 \\
p-HCN-7 & $<$1.05 &  7.9  & $<$0.15 & 5.0 & $<$0.03  & $>$9.7 \\
p-HC$_3$N-8 & $>$0.24 &  7.9 & $\sim$0.73 & 2.0  & $\sim$0.37  &7.3\\
\hline
Z-9 & $<$0.66 & 3.2 	& $<$0.12 & 3.16  & 	$<$0.04 & $>$9.6\\
Z-10  &$<$0.98 & 15.9 & $<$0.08 & 5.01  &	$<$0.02	 & $\gtrsim$10\\
Z-11 & $<$1.02 & 12.6 &$<$0.05 & 2.00  &	$<$0.03	  & $\gtrsim$10 \\
Z-12  & $<$0.62 & 10.0 & $<$0.27& 1.58 &	$<$0.17	 & $>$8.6\\
Z-13  & $<$0.49 & 2.0 & $<$0.03 & 1.58 &	$<$0.02	 & $\gtrsim$10\\
Z-14  & $<$0.57 & 4.0 & $<$0.23 & 2.51 & $<$0.09	  & $>$9.2 \\
\hline\noalign{\smallskip} 	
\end{tabular}
\end{small}
\begin{minipage}{8.5cm}
\small
{{\bf Notes:} Column (2): Size derived as in Sect. \ref{size_par}. 
Column (3) \cite{Emig20} gas mass estimates. The~($^{\star}$) symbol indicates the mass of gas derived as described in \cite{Emig20} Sect.~5.6. 
Column (4): Protostellar luminosity of the SSCs (see Sect.~\ref{lum_derive}). When the v$_{71}$ emission is not detected, we used the the upper limit to the size and the upper limit to the $T_{\rm vib}$ listed in Columns 2 and 5 in Table~\ref{Input_data}, respectively. 
Column (5): The~($^{\star\star}$) symbol indicates the we have estimated the $M_{\rm ZAMS}$ using the 90 GHz continuum emission following \cite{Emig20}. Column~(7): Estimated age of SSCs: for $L_{\rm p*}$/$L_{\rm ZAMS}$$\gtrsim$0.05, $t_{\rm age}$=10$^5$/(1 + $L_{\rm p*}$/$L_{\rm ZAMS}$) while for $L_{\rm p*}$/$L_{\rm ZAMS}$$<$0.05, $t_{\rm age}$$\gtrsim$10$^5$ yr (assuming the timescale of an HII region to be $\sim$10$^5$ yr). 
}
\end{minipage}
\end{table}

We have also identified a clear linear trend ($\rho$$\sim$0.7) between the rotational temperature derived from the HC$_3$N,v$_0$ line with the evolutionary stage of the SSCs, as illustrated in Fig.~\ref{fig_pos_Trot_2}. 
The proto-HC$_3$N show the highest $T_{\rm rot}$ ($\sim$65 K), while the proto-HCN show moderate rotational temperatures ($\sim$45 K). 
Finally, the ZAMS-SSCs show rather low $T_{\rm rot}$ ($\sim$25 K; see also Table~\ref{Params_physical}).
In the earliest phase of star formation of the SSCs, with $t_{\rm age}$ of few 10$^4$ yr (proto-HC$_3$N phase), the radiation of the protostars excites HC$_3$N$^*$ together with HCN$^*$ emission, which increases $T_{\rm rot}$.
This is followed by a more evolved phase (proto-HCN) with $t_{\rm age}$ closer to 10$^5$ yr with moderate $T_{\rm rot}$, likely when the HII regions start to dominate the heating and the radiative feedback in the SSCs, photodissociating the most fragile molecules like HC$_3$N and HNCO (see also \citealt{Rico20, Rico21}). Finally, the ZAMS dominated phase is characterized by rather low $T_{\rm rot}$ and $t_{\rm age}$$\gtrsim$10$^5$ yr. ZAMS-12 and -13 depart from this trend, showing a$T_{\rm rot}$ ($\sim$50 K) closer to that derived for proto-HCN. The higher $T_{\rm rot}$ suggests that they could maybe going through a transition phase between proto-HCN and ZAMS phases. The low excitation temperature of the parent cloud when the SSC reaches the ZAMS stage indicates that the heating by its ultracompact HII regions is mainly restricted to their very close neighborhood.
The inner warm gas heated in the proto-HCN is likely dissociated and only the remaining outer cold parental cloud is observed.

\vspace{-0.4cm}
\subsection{Triggering the SSC formation. Positive feedback}
\label{new_region}

\begin{figure}
 \centering
 \includegraphics[width=0.28\textwidth, angle=180]{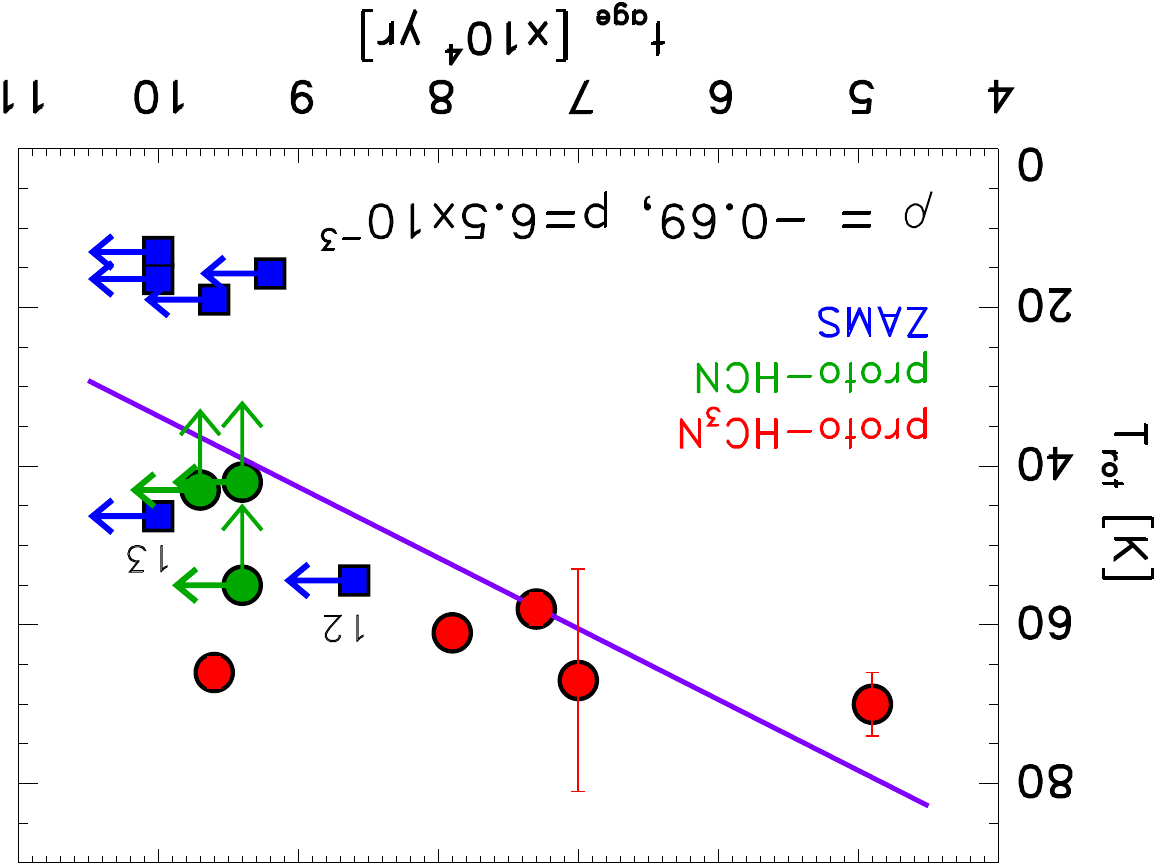}
\vskip-2mm
 \caption{Linear $T_{\rm rot}$-$t_{\rm age}$ relation for the different SSCs. The Spearman's rank correlation coefficient, $\rho$, is derived along with the probability of non-correlation, {\it p} (fit in purple). Same color code as in Fig.~\ref{positions_regs}. } 
\label{fig_pos_Trot_2}
\end{figure}

\begin{figure}
\centering
\includegraphics[width=0.37\textwidth]{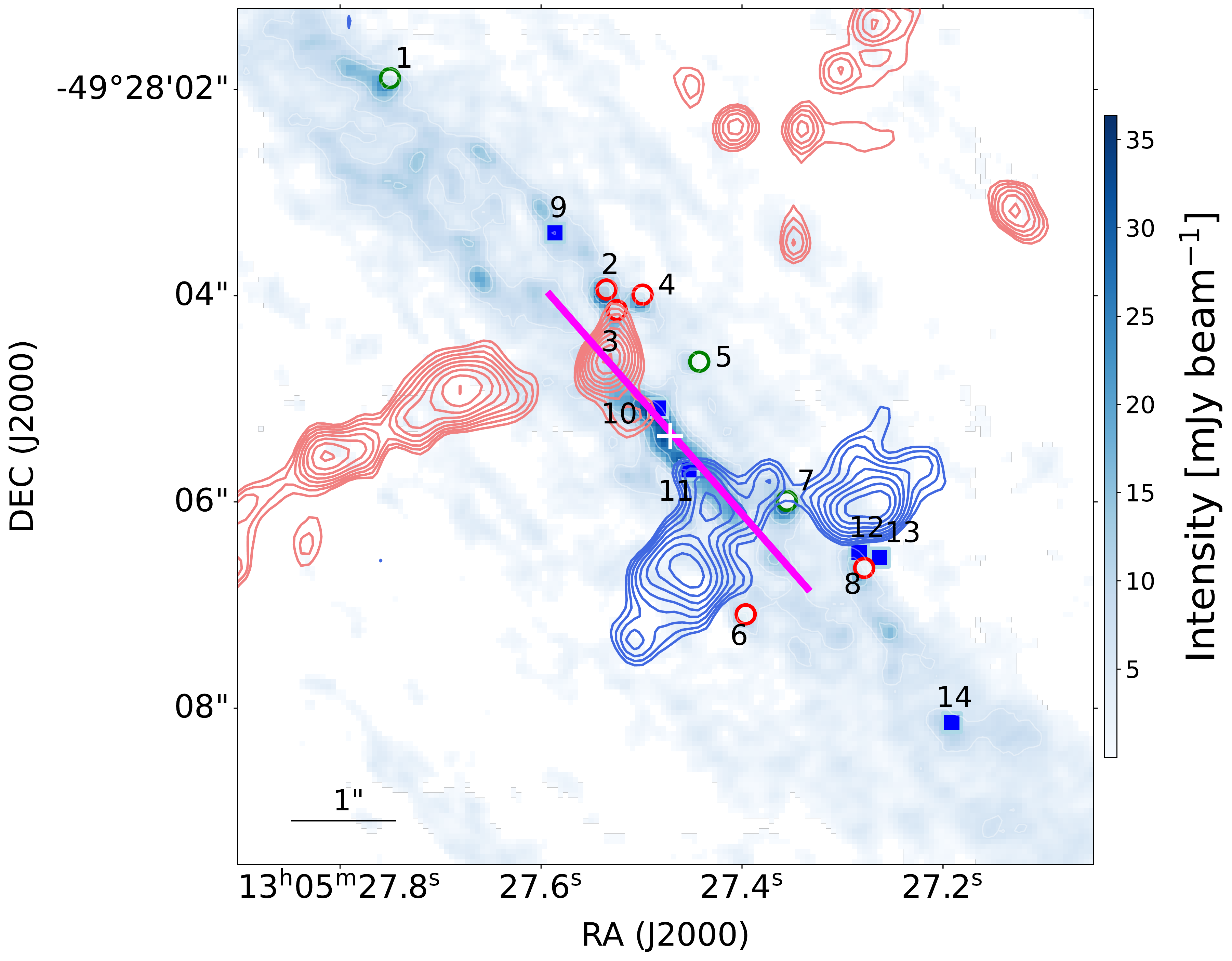}
\vskip-1mm
\caption{Dust emission at $\sim$350 GHz, the CO(3-2) outflow contours and the sources studied in the paper overlaid. The blue and red contours show the approaching and receding sides of the outflow, respectively, with velocity of $\pm$260 km s$^{-1}$ (\citealt{Bolatto21}). The magenta solid line on the disk plane identifies the size of $\pm$2\arcsec\ from the nucleus (see Sect.~\ref{new_region}). The physical scale (1\arcsec$\sim$19 pc) is shown in the lower left corner. Same color code as in Fig.~\ref{positions_regs}. }
\label{outflow_map}
\end{figure}

An inspection of Figs.~\ref{fig_pos_Trot_2} and \ref{outflow_map} shows that the youngest proto-SSCs with ages of 7$\times$10$^4$ yr are concentrated at both sides with respect to the major axis of the galaxy at a projected distance of -1.6\arcsec\ and +1.9\arcsec\ from the nucleus. Adopting an inclination of $\sim$62$^\circ$ for the inner disk (\citealt{Chou07}), we derived a deprojected distance of $\sim$65 and $\sim$80 pc, respectively. \cite{Henkel18} proposed the presence of an inflow of gas in NGC~4945 through a bar from (deprojected) galactocentric radii of 300 pc to $\sim$100 pc toward the nuclear disk, while the central parts ($\lesssim$100 pc) are disturbed by the outflowing gas. Our SSCs fall well within the outflow region, which makes less feasible that the inflow could induce star-formation.
Furthermore, it is remarkable that the proto-HC$_3$N are just bordering the outer edges of the molecular outflow observed in CO by \cite{Bolatto21}, as clearly illustrated in Fig.~\ref{outflow_map}. In the region dominated by the outflow, we also found two out of three proto-HCN (\#5 and \#7), which are probably located outside the molecular outflow but likely associated with it. 
\cite{Bolatto21} investigated the stability of the clumps in the outflow using the virial parameter, $\alpha_{vir}$ finding that most of the gas in the outflow is unlikely to form stars.
From the projected location of basically all the young proto-HC$_3$N, just at the edges of the blue and redshifted lobes of the molecular outflow, we propose that their formation could have been triggered in the shocked gas produced by the expansion of the molecular outflow. 

\cite{Heckman90} derived a dynamical age for the ionized outflow in NGC~4945 to be in the range 3-10 Myr. Even for the molecular phase of the outflow, which shows a more compact distribution ($\sim$70 pc), a dynamical timescale of about 0.3 Myr is derived (\citealt{Bolatto21}). Both dynamical times are larger than the age of the proto-SSCs, making it possible the triggering of the most recent episode of massive star formation in NGC~4945 as seen in HC$_3$N$^*$ and HCN$^*$ by the outflow.

\vspace{-0.4cm}
\subsection{The evolution of the star formation in NGC~4945 and NGC~253. Sequential SSC formation.}

While the typical age of the proto-SSCs in NGC~4945 are about $>$0.8$\times$10$^5$ ($\pm<$0.2$\times$10$^5$) yr, those in NGC~253 are younger, with $t_{\rm age}$$\sim$0.2-0.8~$\times$10$^5$ yr. The derived ages for the SSCs in NGC~4945 indicate that the burst of star formation in this galaxy seems to be more evolved than in NCG~253, in agreement with the age of the burst in each galaxy (i.e., $\sim$1 My for NGC~253 and $\gtrsim$5 Myr for NGC~4945; \citealt{Leroy18, Emig20}).
Also the mean gas mass reservoir available for star formation in the proto-SSCs in NGC~253 ($\sim$1.7$\times$10$^5$~$M_{\sun}$) is larger, by at least a factor of 2, than in the NGC~4945 (7.5$\times$10$^4$~$M_{\sun}$), also consistent with the different ages of the starburst in both galaxies.
In fact, the gas mass reservoir in NGC~4945 is quite similar in both (proto and ZAMS) phases (see Table~\ref{Params_physical}), of about 7$\times$10$^4$~$M_{\sun}$, indicating that the burst of star formation in NGC~4945 is likely reaching its end, as in NGC~253, which also shares similar amount of dense gas (HNCO/CS) with NGC~4945 (\citealt{Martin09}).

In fact, the different origins of the youngest proto-SSCs in both galaxies can be understood in the context of the scenario of feedback and sequential SSC formation. While the most evolved SSCs in NGC~253 are mainly located closer to the center of the galaxy and the younger proto-SSCs in the outer parts (`inside-out' scenario), in NGC~4945 this trend is the opposite, with most of the youngest proto-SSCs found toward the center. 
The SSCs age and spatial distribution in NGC~253 seem to suggest that the SSCs formation have been triggered by the overpressure produced by hot gas generated by the SN explosions from an early star formation episode in the center of the galaxy (\citealt{Rico20}). In the case of more evolved starburst as  NGC~4945, the well developed and strong molecular outflow, from the previous episode of star formation, induced the new generation of SSCs by positive feedback.
Recently, a face-on (deprojected) view has been proposed in NGC~253 by \cite{Levy22}, which still confirms the sequential star formation scenario of SSCs. Furthermore, according to the model shown by \citet[right panel in their Fig.~6]{Bolatto21}, the results obtained considering the deprojection of NGC~4945 would be still valid in a face-on view scenario.

\vspace{-0.5cm}
\section{Conclusions}

We present ALMA imaging of vibrationally excited HC$_3$N and HCN emissions to study the earliest phase of the formation and evolution of SSCs in the composite SB/AGN galaxy NGC~4945.
Our main results can be summarized as follows: 

\begin{itemize}
\item We have identified 14 forming SSCs in the center of NGC~4945: 8 of them show HCN vibrational excited emission (HCN$^*$), 5 of which also present HC$_3$N vibrationally excited emission (HC$_3$N$^*$). The other 6 SSCs do not show any vibrationally excited emissions in spite of their large stellar masses. From the LTE analysis of the HC$_3$N$^*$ lines, we derived both rotational ($T_{\rm rot}$) and vibrational ($T_{\rm vib}$) temperatures of the molecular gas associated with the SSCs;

\item We found that the 8 SSCs detected in vibrationally excited emission are in the proto phase, characterized by a mean age of $\sim$7$\times$10$^4$~yr, in contrast with the 6 more evolved SSCs, which are in the ZAMS phase with ages~$\gtrsim$10$^5$ yr;

\item
We identified a clear trend between $T_{\rm rot}$ from HC$_3$N with the different stages of evolution of the SSCs. 
The earliest phase of SSC formation with $t_{\rm age}$$\lesssim$7$\times$10$^4$ yr is traced by HC$_3$N$^*$ and HCN$^*$ vibrationally excited emissions and high $T_{\rm rot}$, then followed by a more evolved phase traced by the detection of only HCN vibrationally excited emission, with moderate $T_{\rm rot}$ and $t_{\rm age}$ closer to 10$^5$~yr, when HII regions start to dominate the heating and the radiative feedback (photodissociation). Finally, the ZAMS phase is characterized by rather low $T_{\rm rot}$, and $t_{\rm age}$$\gtrsim$10$^5$ yr, without detection of the vibrationally excited emissions; 

\item We found that the youngest proto-SSCs are located at the edges of the outflow (\citealt{Bolatto21}), suggesting that the outflow has triggered the formation of the youngest proto-SSCs (positive feedback) in the nearby shocked regions;
 
\item SSCs in NGC~4945 are found to be in a more evolved stage than those in NGC~253, consistent with the sequential SSC formation scenario. While in NGC~253 SSCs seem to be formed by the overpressure produced by the SN explosions of the previous generation of stars, the youngest generation of SSCs in NGC~4945 have been triggered in the shocked regions produced by the well developed molecular outflow. 
\end{itemize}

\vspace{-0.7cm}
\section*{Acknowledgements}

\vskip-1mm
EB acknowledges the Mar\'{\i}a Zambrano program of the Spanish Ministerio de Universidades funded by the Next Generation European Union and is also partly supported by grant RTI2018-096188-B-I00 funded by MCIN/AEI/10.13039/501100011033.
EB and JMP acknowledge the support from grant No. PID2019-105552RB-C41 by the Spanish Ministry of Science and Innovation/State Agency of Research MCIN/AEI/10.13039/501100011033 and by ``ERDF A way of making Europe'', and JMP by Comunidad de Madrid under Grant P2018/NMT-4291. This paper makes use of the following ALMA data: ADS/JAO.ALMA\#2016.1.01135.S and ADS/JAO.ALMA\#2018.1.01236.S. ALMA is a partnership of ESO (representing its member states), NSF (USA) and NINS (Japan), together with NRC (Canada), MOST and ASIAA (Taiwan), and KASI (Republic of Korea), in cooperation with the Republic of Chile. The Joint ALMA Observatory is operated by ESO, AUI/NRAO and~NAOJ.

\vspace{-0.7cm}
\section*{Data availability}

The data underlying this article were accessed from the Alma Science Archive \url{http://almascience.eso.org/asax/} with the corresponding project identifiers listed in Section~2. The derived data generated in this research will be shared on reasonable request to the corresponding author.

\vspace{-0.7cm}
\bibliographystyle{mnras} 
\bibliography{biblio}     

\begin{thebibliography}{}
\makeatletter
\relax
\def\mn@urlcharsother{\let\do\@makeother \do\$\do\&\do\#\do\^\do\_\do\%\do\~}
\def\mn@doi{\begingroup\mn@urlcharsother \@ifnextchar [ {\mn@doi@}
  {\mn@doi@[]}}
\def\mn@doi@[#1]#2{\def\@tempa{#1}\ifx\@tempa\@empty \href
  {http://dx.doi.org/#2} {doi:#2}\else \href {http://dx.doi.org/#2} {#1}\fi
  \endgroup}
\def\mn@eprint#1#2{\mn@eprint@#1:#2::\@nil}
\def\mn@eprint@arXiv#1{\href {http://arxiv.org/abs/#1} {{\tt arXiv:#1}}}
\def\mn@eprint@dblp#1{\href {http://dblp.uni-trier.de/rec/bibtex/#1.xml}
  {dblp:#1}}
\def\mn@eprint@#1:#2:#3:#4\@nil{\def\@tempa {#1}\def\@tempb {#2}\def\@tempc
  {#3}\ifx \@tempc \@empty \let \@tempc \@tempb \let \@tempb \@tempa \fi \ifx
  \@tempb \@empty \def\@tempb {arXiv}\fi \@ifundefined
  {mn@eprint@\@tempb}{\@tempb:\@tempc}{\expandafter \expandafter \csname
  mn@eprint@\@tempb\endcsname \expandafter{\@tempc}}}

\bibitem[\protect\citeauthoryear{{Bellocchi} et~al.,}{{Bellocchi}
  et~al.}{2020}]{Bellocchi20}
{Bellocchi} E.,  et~al., 2020, \mn@doi [\aap] {10.1051/0004-6361/202037782},
  \href {https://ui.adsabs.harvard.edu/abs/2020A&A...642A.166B} {642, A166}

\bibitem[\protect\citeauthoryear{{Bolatto} et~al.,}{{Bolatto}
  et~al.}{2013}]{Bolatto13}
{Bolatto} A.~D.,  et~al., 2013, \mn@doi [\nat] {10.1038/nature12351}, \href
  {https://ui.adsabs.harvard.edu/abs/2013Natur.499..450B} {499, 450}

\bibitem[\protect\citeauthoryear{{Bolatto} et~al.,}{{Bolatto}
  et~al.}{2021}]{Bolatto21}
{Bolatto} A.~D.,  et~al., 2021, \mn@doi [\apj] {10.3847/1538-4357/ac2c08},
  \href {https://ui.adsabs.harvard.edu/abs/2021ApJ...923...83B} {923, 83}

\bibitem[\protect\citeauthoryear{{Chou} et~al.,}{{Chou} et~al.}{2007}]{Chou07}
{Chou} R. C.~Y.,  et~al., 2007, \mn@doi [\apj] {10.1086/521351}, \href
  {https://ui.adsabs.harvard.edu/abs/2007ApJ...670..116C} {670, 116}

\bibitem[\protect\citeauthoryear{{Dahlem}, {Golla}, {Whiteoak}, {Wielebinski},
  {Huettemeister}  \& {Henkel}}{{Dahlem} et~al.}{1993}]{Dahlem93}
{Dahlem} M.,  {Golla} G.,  {Whiteoak} J.~B.,  {Wielebinski} R.,
  {Huettemeister} S.,   {Henkel} C.,  1993, \aap, \href
  {https://ui.adsabs.harvard.edu/abs/1993A&A...270...29D} {270, 29}

\bibitem[\protect\citeauthoryear{{Emig} et~al.,}{{Emig} et~al.}{2020}]{Emig20}
{Emig} K.~L.,  et~al., 2020, \mn@doi [\apj] {10.3847/1538-4357/abb67d}, \href
  {https://ui.adsabs.harvard.edu/abs/2020ApJ...903...50E} {903, 50}

\bibitem[\protect\citeauthoryear{{Gonz{\'a}lez-Alfonso} \&
  {Sakamoto}}{{Gonz{\'a}lez-Alfonso} \& {Sakamoto}}{2019}]{GA19}
{Gonz{\'a}lez-Alfonso} E.,  {Sakamoto} K.,  2019, \mn@doi [\apj]
  {10.3847/1538-4357/ab3a32}, \href
  {https://ui.adsabs.harvard.edu/abs/2019ApJ...882..153G} {882, 153}

\bibitem[\protect\citeauthoryear{{Heckman}, {Armus}  \& {Miley}}{{Heckman}
  et~al.}{1990}]{Heckman90}
{Heckman} T.~M.,  {Armus} L.,   {Miley} G.~K.,  1990, \mn@doi [\apjs]
  {10.1086/191522}, \href
  {https://ui.adsabs.harvard.edu/abs/1990ApJS...74..833H} {74, 833}

\bibitem[\protect\citeauthoryear{{Henkel} et~al.,}{{Henkel}
  et~al.}{2018}]{Henkel18}
{Henkel} C.,  et~al., 2018, \mn@doi [\aap] {10.1051/0004-6361/201732174}, \href
  {https://ui.adsabs.harvard.edu/abs/2018A&A...615A.155H} {615, A155}

\bibitem[\protect\citeauthoryear{{Herrera}, {Boulanger}  \&
  {Nesvadba}}{{Herrera} et~al.}{2011}]{Herrera11}
{Herrera} C.~N.,  {Boulanger} F.,   {Nesvadba} N.~P.~H.,  2011, \mn@doi [\aap]
  {10.1051/0004-6361/201117324}, \href
  {https://ui.adsabs.harvard.edu/abs/2011A&A...534A.138H} {534, A138}

\bibitem[\protect\citeauthoryear{{Krieger} et~al.,}{{Krieger}
  et~al.}{2019}]{Krieger19}
{Krieger} N.,  et~al., 2019, \mn@doi [\apj] {10.3847/1538-4357/ab2d9c}, \href
  {https://ui.adsabs.harvard.edu/abs/2019ApJ...881...43K} {881, 43}

\bibitem[\protect\citeauthoryear{{Leroy} et~al.,}{{Leroy}
  et~al.}{2018}]{Leroy18}
{Leroy} A.~K.,  et~al., 2018, \mn@doi [\apj] {10.3847/1538-4357/aaecd1}, \href
  {https://ui.adsabs.harvard.edu/abs/2018ApJ...869..126L} {869, 126}

\bibitem[\protect\citeauthoryear{{Levy} et~al.,}{{Levy} et~al.}{2022}]{Levy22}
{Levy} R.~C.,  et~al., 2022, \mn@doi [\apj] {10.3847/1538-4357/ac7b7a}, \href
  {https://ui.adsabs.harvard.edu/abs/2022ApJ...935...19L} {935, 19}

\bibitem[\protect\citeauthoryear{{Mart{\'\i}n}, {Mart{\'\i}n-Pintado}  \&
  {Mauersberger}}{{Mart{\'\i}n} et~al.}{2009}]{Martin09}
{Mart{\'\i}n} S.,  {Mart{\'\i}n-Pintado} J.,   {Mauersberger} R.,  2009,
  \mn@doi [\apj] {10.1088/0004-637X/694/1/610}, \href
  {https://ui.adsabs.harvard.edu/abs/2009ApJ...694..610M} {694, 610}

\bibitem[\protect\citeauthoryear{{Mart{\'\i}n}, {Mart{\'\i}n-Pintado},
  {Blanco-S{\'a}nchez}, {Rivilla}, {Rodr{\'\i}guez-Franco}  \&
  {Rico-Villas}}{{Mart{\'\i}n} et~al.}{2019}]{Martin19}
{Mart{\'\i}n} S.,  {Mart{\'\i}n-Pintado} J.,  {Blanco-S{\'a}nchez} C.,
  {Rivilla} V.~M.,  {Rodr{\'\i}guez-Franco} A.,   {Rico-Villas} F.,  2019,
  \mn@doi [\aap] {10.1051/0004-6361/201936144}, \href
  {https://ui.adsabs.harvard.edu/abs/2019A&A...631A.159M} {631, A159}

\bibitem[\protect\citeauthoryear{{Mingozzi} et~al.,}{{Mingozzi}
  et~al.}{2019}]{Mingozzi19}
{Mingozzi} M.,  et~al., 2019, \mn@doi [\aap] {10.1051/0004-6361/201834372},
  \href {https://ui.adsabs.harvard.edu/abs/2019A&A...622A.146M} {622, A146}

\bibitem[\protect\citeauthoryear{{M{\"u}ller}, {Thorwirth}, {Roth}  \&
  {Winnewisser}}{{M{\"u}ller} et~al.}{2001}]{Muller01}
{M{\"u}ller} H.~S.~P.,  {Thorwirth} S.,  {Roth} D.~A.,   {Winnewisser} G.,
  2001, \mn@doi [\aap] {10.1051/0004-6361:20010367}, \href
  {https://ui.adsabs.harvard.edu/abs/2001A&A...370L..49M} {370, L49}

\bibitem[\protect\citeauthoryear{{Pickett}, {Poynter}, {Cohen}, {Delitsky},
  {Pearson}  \& {M{\"u}ller}}{{Pickett} et~al.}{1998}]{Pickett98}
{Pickett} H.~M.,  {Poynter} R.~L.,  {Cohen} E.~A.,  {Delitsky} M.~L.,
  {Pearson} J.~C.,   {M{\"u}ller} H.~S.~P.,  1998, \mn@doi [\jqsrt]
  {10.1016/S0022-4073(98)00091-0}, \href
  {https://ui.adsabs.harvard.edu/abs/1998JQSRT..60..883P} {60, 883}

\bibitem[\protect\citeauthoryear{{Rico-Villas}, {Mart{\'\i}n-Pintado},
  {Gonz{\'a}lez-Alfonso}, {Mart{\'\i}n}  \& {Rivilla}}{{Rico-Villas}
  et~al.}{2020}]{Rico20}
{Rico-Villas} F.,  {Mart{\'\i}n-Pintado} J.,  {Gonz{\'a}lez-Alfonso} E.,
  {Mart{\'\i}n} S.,   {Rivilla} V.~M.,  2020, \mn@doi [\mnras]
  {10.1093/mnras/stz3347}, \href
  {https://ui.adsabs.harvard.edu/abs/2020MNRAS.491.4573R} {491, 4573}

\bibitem[\protect\citeauthoryear{{Rico-Villas}, {Mart{\'\i}n-Pintado},
  {Gonz{\'a}lez-Alfonso}, {Rivilla}, {Mart{\'\i}n}, {Garc{\'\i}a-Burillo},
  {Jim{\'e}nez-Serra}  \& {S{\'a}nchez-Garc{\'\i}a}}{{Rico-Villas}
  et~al.}{2021}]{Rico21}
{Rico-Villas} F.,  {Mart{\'\i}n-Pintado} J.,  {Gonz{\'a}lez-Alfonso} E.,
  {Rivilla} V.~M.,  {Mart{\'\i}n} S.,  {Garc{\'\i}a-Burillo} S.,
  {Jim{\'e}nez-Serra} I.,   {S{\'a}nchez-Garc{\'\i}a} M.,  2021, \mn@doi
  [\mnras] {10.1093/mnras/stab197}, \href
  {https://ui.adsabs.harvard.edu/abs/2021MNRAS.502.3021R} {502, 3021}

\bibitem[\protect\citeauthoryear{{Rico-Villas}, {Gonz{\'a}lez-Alfonso},
  {Mart{\'\i}n-Pintado}, {Rivilla}  \& {Mart{\'\i}n}}{{Rico-Villas}
  et~al.}{2022}]{Rico22}
{Rico-Villas} F.,  {Gonz{\'a}lez-Alfonso} E.,  {Mart{\'\i}n-Pintado} J.,
  {Rivilla} V.~M.,   {Mart{\'\i}n} S.,  2022, \mn@doi [\mnras]
  {10.1093/mnras/stac2260}, \href
  {https://ui.adsabs.harvard.edu/abs/2022MNRAS.tmp.2149R} {}

\bibitem[\protect\citeauthoryear{{Schurch}, {Roberts}  \& {Warwick}}{{Schurch}
  et~al.}{2002}]{Schurch02}
{Schurch} N.~J.,  {Roberts} T.~P.,   {Warwick} R.~S.,  2002, \mn@doi [\mnras]
  {10.1046/j.1365-8711.2002.05585.x}, \href
  {https://ui.adsabs.harvard.edu/abs/2002MNRAS.335..241S} {335, 241}

\bibitem[\protect\citeauthoryear{{Strickland}, {Heckman}, {Weaver}, {Hoopes}
  \& {Dahlem}}{{Strickland} et~al.}{2002}]{Strickland02}
{Strickland} D.~K.,  {Heckman} T.~M.,  {Weaver} K.~A.,  {Hoopes} C.~G.,
  {Dahlem} M.,  2002, \mn@doi [\apj] {10.1086/338889}, \href
  {https://ui.adsabs.harvard.edu/abs/2002ApJ...568..689S} {568, 689}

\bibitem[\protect\citeauthoryear{{Wall}, {Mac Low}, {McMillan}, {Klessen},
  {Portegies Zwart}  \& {Pellegrino}}{{Wall} et~al.}{2020}]{Wall20}
{Wall} J.~E.,  {Mac Low} M.-M.,  {McMillan} S. L.~W.,  {Klessen} R.~S.,
  {Portegies Zwart} S.,   {Pellegrino} A.,  2020, \mn@doi [\apj]
  {10.3847/1538-4357/abc011}, \href
  {https://ui.adsabs.harvard.edu/abs/2020ApJ...904..192W} {904, 192}

\bibitem[\protect\citeauthoryear{{Westmoquette}, {Smith}  \&
  {Gallagher}}{{Westmoquette} et~al.}{2011}]{Westmoquette11}
{Westmoquette} M.~S.,  {Smith} L.~J.,   {Gallagher} J.~S. I.,  2011, \mn@doi
  [\mnras] {10.1111/j.1365-2966.2011.18675.x}, \href
  {https://ui.adsabs.harvard.edu/abs/2011MNRAS.414.3719W} {414, 3719}

\makeatother
\end{thebibliography}

\bsp	
\label{lastpage}
\end{document}